\documentclass[11pt,a4paper]{article}

\usepackage{graphicx}

\title{An Automated Approach
for $q\bar{q}\rightarrow b\bar{b}b\bar{b}$
at Next-to-Leading Order QCD}

\author{%
T.~Reiter\footnote{%
	Nikhef,
	Science Park 105,
	1098~XG~Amsterdam,
	The~Netherlands}
}

\date{12~February~2009}
\begin{document}

\maketitle

\begin{abstract}
The search for the Higgs boson and for physics beyond the
Standard Model are the major motivations behind the LHC
experiment. In many scenarios the success of the experiment
depends on the knowledge of signal and background event rates
at least at one-loop precision.
We present the approach of the \texttt{GOLEM} collaboration to build
a highly automated framework for the calculation of matrix elements
at the one-loop level, which is based on the evaluation of Feynman
diagrams. Part of this effort is an open-source library for
the numerical evaluation of tensor integrals.
As an application, some results for the process
$pp\rightarrow b\bar{b}b\bar{b}$ calculated with this method are
presented.
\end{abstract}

\paragraph{PACS} %
	12{.}38{.}Bx, 
	13{.}85{.}Hd 


\section{Introduction}
The Standard Model of Particle Physics has been tested by previous
collider experiments to a very high precision~\cite{Collaboration:2008ub}.
Despite its great success, many fundamental questions such as the nature
of dark matter cannot be addressed within the Standard Model.
Two main goals of the next collider experiment, the LHC, are
therefore the discovery (or exclusion) of a Standard Model Higgs boson
and the measurement of any new particles accessible by the collider
energy~\cite{:1999fr}.
Due to the purely hadronic initial state of the collisions one expects
the interactions mainly to be governed by QCD.
A precise understanding of both the signal and the background will
be crucial for most Higgs discovery channels and for the discrimination
of different scenarios beyond the Standard Model. For many processes
a Leading Order (LO) approximation will not suffice and has to be
amended by higher order corrections.

The calculation of matrix elements at LO has become an automated routine
for which many computer programs are available~\cite{%
Kilian:2007gr,%
Moretti:2001zz,Maltoni:2002qb,Krauss:2001iv,Boos:2004kh,Mangano:2002ea}.
Next to Leading Order (NLO) calculations, however, have never reached
this level of automation. Especially in the case of many particle final
states ($\geq 3$ final state particle) automatisation is not
straightforward as one easily hits the limits of current computer technology.
A full NLO calculation in QCD consists of a
$2\rightarrow n$ particle tree-level contribution (LO),
the real emission of an extra parton ($2\rightarrow n+1$, tree-level)
and virtual corrections ($2\rightarrow n$, one-loop diagrams).
Both real and virtual corrections can contain infrared divergences
which only cancel in the sum of both contributions. These divergences
can be dealt with by subtraction methods~\cite{Catani:1996jh,Catani:2002hc}
which have also become available as automated
implementations~\cite{Gleisberg:2007md,Seymour:2008mu,%
Hasegawa:2008ae,Frederix:2008hu}.
The only missing ingredient for a full automation of NLO calculations
are the virtual corrections.
Although many different methods have been
proposed~\cite{Berger:2008sj,Catani:2008xa,Giele:2008bc,Ellis:2008ir,%
Britto:2008vq,Ossola:2006us,Denner:2005nn,Bern:2007dw}
no fully
automated implementation has been made available yet.
The very limited number of results found in the literature for processes
with four final state particles~\cite{Bredenstein:2008zb,Binoth:2008gx,%
Berger:2008sz,Ellis:2008qc,Berger:2009zg,Reiter:2009kb} underlines the
importance of automatisation in the context of one-loop calculations.

The \texttt{GOLEM} collaboration focuses on the development of such
an automatised tool for one-loop matrix element calculations\footnote{%
\texttt{GOLEM} stands for General One Loop Evaluator for Matrix elements}.
We have applied the \texttt{GOLEM} method to calculate the
QCD one-loop corrections of the process
$u\bar{u}\rightarrow b\bar{b}b\bar{b}$, which is a subprocess of
$pp\rightarrow b\bar{b}b\bar{b}$. This process is a particular important
background in MSSM Higgs searches at large values of $\tan\beta$, where
one of the Higgs bosons decays predominantly into $b\bar{b}$ pairs.

\section{The \texttt{GOLEM} Approach}
Our approach is based on the calculation of Feynman diagrams.
We generate the diagrams and their corresponding algebraic expressions
using \texttt{QGraf}~\cite{Nogueira:1991ex} and project them on a
colour and helicity basis. The integration over the momentum of the
virtual particle introduces tensor integrals of the form
\begin{equation}
I^{n;\mu_1,\ldots,\mu_r}_N=\int\!\!\frac{\mathrm{d}^nk}{i\pi^{n/2}}%
\frac{k^{\mu_1}\cdots k^{\mu_r}}{\prod_{j=1}^N\left[(k+r_j)^2-m_j^2+i\delta%
\right]}
\end{equation}
These integrals are reduced to a basis of scalar integrals by
the method described in~\cite{Binoth:1999sp,Binoth:2005ff}; we
use two independent implementations where
\begin{enumerate}
\item[a)] the tensor reduction is carried out on a symbolical level and
the amplitude is represented in terms of Mandelstam variables,
Levi-Civita tensors
and the commonly used standard basis of scalar one-loop integrals
\begin{displaymath}
\mathcal{M}_{\{\lambda\},c}=C_{\mathrm{box}}I_4^n
+C_{\mathrm{tri}}I_3^n+C_{\mathrm{bub}}I_2^n+C_{\mathrm{tad}}I_1^n.
\end{displaymath}
The algebraic reduction and further simplifications 
are achieved using \texttt{Form}~\cite{Vermaseren:2000nd} and~Maple.
\item[b)] a form factor representation is introduced for the tensor
integrals and the tensor reduction is delayed until the numerical evaluation.
The form factors are implemented in the Fortran library \texttt{golem95}
which is described in~\cite{Binoth:2008uq}. The amplitude is represented
in terms of Mandelstam variables,
spinor bi-products $\bar{u}(p_i)(1\pm\gamma_5)u(p_j)$ and an extended
basis of one-loop integrals that allows for Feynman parameters in the
numerator.
\end{enumerate}
In both strategies, Gram determinants in the denominator are either
avoided or to some extent cancelled explicitly
on the symbolical level, thus resulting
in a numerically stable implementation of the matrix element.

The results presented below have been obtained with implementation~b)
in which the user starts from a very minimal process description. All
necessary files are generated by a \texttt{Python} script. The matrix
element is obtained as a set of \texttt{Fortran90} files which are
compiled on the target system, in our case the
ECDF cluster~\cite{ECDF}.

The direct integration of a one-loop matrix element over phase space with
an adaptive Monte Carlo (MC) has two major disadvantages. The evaluation
time of an NLO matrix element is considerable larger than that of a
LO matrix element. Hence one should try to avoid unnecessary calls of the
NLO matrix elements. The second problem are phase space regions where
the chosen integral basis becomes linearly dependent and lead to numerical
fluctuations. When these fluctuations reach the order of magnitude of the
precision goal of the adaptive MC, the MC program tends to overestimate
these phase space regions which can lead to numerical instabilities.

We avoid these problems by performing the phase space integration
as a reweighting of unweighted LO Monte Carlo events.
We use \texttt{WHIZARD}~\cite{Kilian:2007gr}
an adaptive Monte Carlo integrator to obtain a list of unweighted
LO events $\{p\}_i$ such that an observable $O$ can be written as
\begin{equation}
\langle O\rangle_{\mathrm{LO}}\equiv\int\mathrm{d}\Phi(\{p\})
\vert\mathcal{M}_{\mathrm{LO}}\vert^2O(\{p\})=
\lim_{N\rightarrow\infty}
\frac{\sigma_{\mathrm{LO}}}{N}\sum_{i=1}^NO(\{p\}_i)
\end{equation}
where the limit is understood in the statistical sense as a limit on the
variance of the MC sum. The observable at one-loop precision is obtained
as
\begin{equation}
\langle O\rangle_{\mathrm{one-loop}}=
\lim_{N\rightarrow\infty}
\frac{\sigma_{\mathrm{LO}}}{N}\sum_{i=1}^NK(\{p\})O(\{p\}_i)
\end{equation}
with the local $K$-factor
\begin{equation}\label{eq:local-K}
K(\{p\})=\frac{%
\mathcal{M}_{\mathrm{LO}}^\dagger\cdot
(\mathcal{M}_{\mathrm{LO}}+\mathcal{M}_{\mathrm{virt}}
+\mathbf{I}\cdot \mathcal{M}_{\mathrm{LO}})}{%
\vert\mathcal{M}_{\mathrm{LO}}\vert^2}
\end{equation}
The matrix elements $\mathcal{M}$ are understood as vectors in a
given colour basis and $\mathbf{I}$ is the insertion operator
as defined in~\cite{Catani:1996jh}, which ensures that after
UV-renormalisation all poles in $1/(n-4)$ cancel. The reweighting
can be understood as importance sampling with the probability
density
\begin{equation}
w(\{p\})\propto\frac{1}{\sigma_{\mathrm{LO}}}
\frac{\mathrm{d}\sigma_{\mathrm{LO}}(\{p\})}{\mathrm{d}\Phi(\{p\})}.
\end{equation}

It should be emphasized that the definition of the
local $K$-factor in Eq.~(\ref{eq:local-K})
does not contain any real emission contributions
and therefore the results below lack any physical interpretation.

\section{Results for $u\bar{u}\rightarrow b\bar{b}b\bar{b}$}
The results for the virtual correction of the
process~$u\bar{u}\rightarrow b\bar{b}b\bar{b}$ have been obtained
with $n_f=5$ massless quark flavours and for $p_T>30\,\mathrm{GeV}$,
a rapidity cut of $\eta<2{.}5\,\mathrm{GeV}$ and a separation cut
of $\Delta R=\sqrt{(\Delta\Phi)^2+(\Delta\eta)^2}>0{.}4$. The centre of
mass energy is 14\,TeV. In this
data set we choose the scales as $\mu_F=\mu_R=\sum_{i=1}^4p_{T,i}/4$
and fold with the CTEQ6{.}5 parton distribution functions~\cite{Tung:2006tb}.
We work with the modifications to the dipoles and insertion operators
as proposed in~\cite{Nagy:2003tz}; for the distributions below we have
evaluated the insertion operator at $\alpha_{\mathrm{N}}=0{.}1$.
For this specific amplitude this modification 
induces a shift of the insertion operator of
\begin{equation}
\mathbf{I}_{\mathrm{unmodified}}-\mathbf{I}(\alpha_{\mathrm{N}})=
\frac{\alpha_s(\mu)}{2\pi}\cdot 6\,C_F \left[
\ln^2\alpha_{\mathrm{N}}
-\frac32\left(\alpha_{\mathrm{N}}-1-\ln\alpha_{\mathrm{N}}\right)
\right].
\end{equation}
On the ECDF cluster our code achieved a performance of
8{.}9\,s (17{.}6\,s) per phase space point and
node\footnote{Xeon\,5450 (quad-core), 3\,GHz} in double (quadruple)
precision.
The above timings suggest that an evaluation of the whole integration
in quadruple precision is too costly to be practical and should be the
last resort if double precision is not sufficient. The runtime will
be improved further through code optimisations in the next version of
the \texttt{GOLEM} code.

\begin{figure}[hbt]
\begin{center}
\includegraphics[width=0.9\textwidth]{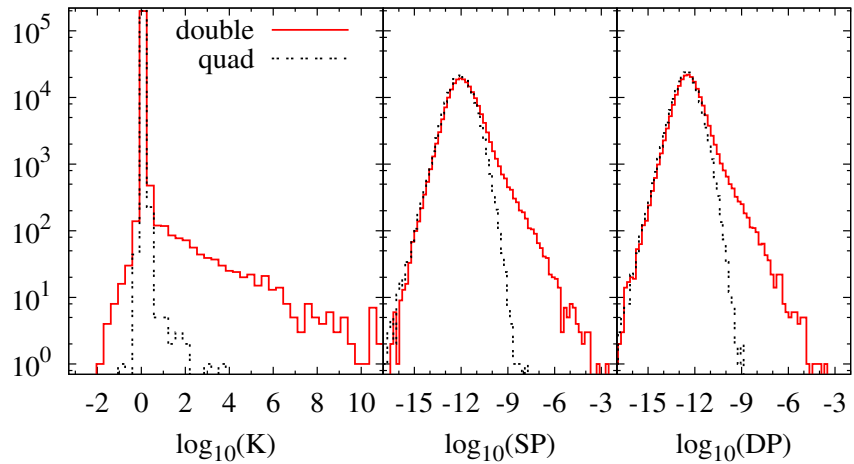}
\end{center}
\caption{Comparison between evaluations of the matrix element for
200{,}000 MC events in double precision and quadruple precision.
Shown are the distributions of the values of the local $K$-factor
as defined in Eq.~(\ref{eq:local-K}) (left) and the single $1/(n-4)$ (middle)
and double $1/(n-4)^2$ (right) pole of the matrix element.}
\label{fig:acc-hist}
\end{figure}

Figure~\ref{fig:acc-hist} shows the distribution of the values for the
local $K$-factor, the single pole and the double pole of an amplitude
for 200{,}000 randomly chosen points in both double and quadruple precision.
The distribution of the the $K$-factor shows that the values are sharply
peaked around the the integrated result of $\mathcal{O}(1)$ and
in double precision a small
fraction of less than a percent of the points stretches out to atypically
large values. It is clear that in a data set of order $10^6$ MC events
a single outlier of that magnitude is already enough
to tamper with the result. On the other hand, since the number of points
in doubt is very small, an a posteriori test is enough and it suffices
to re-evaluate those points at a higher precision. As possible test
criteria we studied a cut on the $K$-factor, on the coefficient of the
single pole and on the double pole; if the magnitude of the double
precision result exceeds the cut the point is evaluated at higher precision.

\begin{figure}[htbp]
\begin{center}
\includegraphics[width=0.9\textwidth]{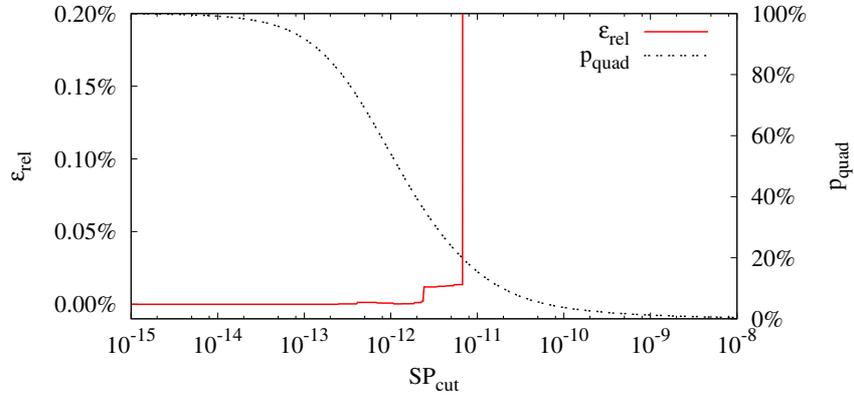}
\end{center}
\caption{Relative error on the integrated result versus a
cut~$\mathrm{SP}_{\mathrm{cut}}$ on the coefficient of the single pole.
If the cancellation of the single pole in double precision
is worse than $\mathrm{SP}_{\mathrm{cut}}$ the data point is evaluated
in quadruple precision. The dashed curve indicates the percentage of points
that fail the test and need re-evaluation (right $y$-axis).}
\label{fig:s-acc}
\end{figure}

\begin{figure}[htbp]
\begin{center}
\includegraphics[width=0.9\textwidth]{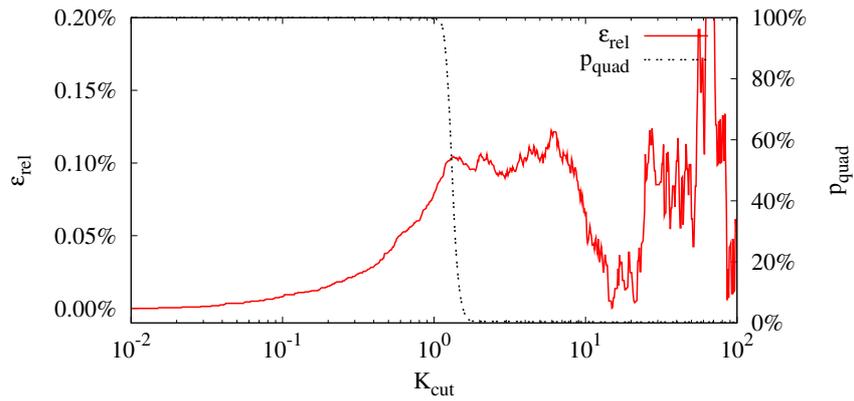}
\end{center}
\caption{Relative error as in Figure~\ref{fig:s-acc} but with a
cut on the local $K$-factor.}
\label{fig:k-acc}
\end{figure}

We have taken the same sample of 200{,}000 points to study the influence
of the cut parameter on the integrated result. Figure~\ref{fig:s-acc}
shows the relative error $\varepsilon_{\mathrm{rel}}=%
\vert\sigma(\mathrm{SP}_{\mathrm{cut}})-\sigma(0)\vert/\sigma(0)$
on the cross-section versus the cut parameter~$\mathrm{SP}_{\mathrm{cut}}$.
The steep increase of the error indicates an outlier in the $K$-factor
that is not reflected in the cancellation of the pole. A very similar
picture emerges for a cut on the double pole (not shown). This lack of
correlation between the pole cancellation and error on the integral
can be circumvented by imposing a test on the local $K$-factor as
shown in Figure~\ref{fig:k-acc}. In the region between
$2\leq K_{\mathrm{cut}}\leq5$ a relative error of $\approx0{.}10\%$ is
achieved while only $0{.}5\%$ of the phase space points need to be
evaluated at a higher precision.\footnote{The downwards trend at
$K_{\mathrm{cut}}\approx10$ is a statistical fluctuation.}

\begin{figure}[htbp]
\begin{center}
\includegraphics[width=0.45\textwidth]{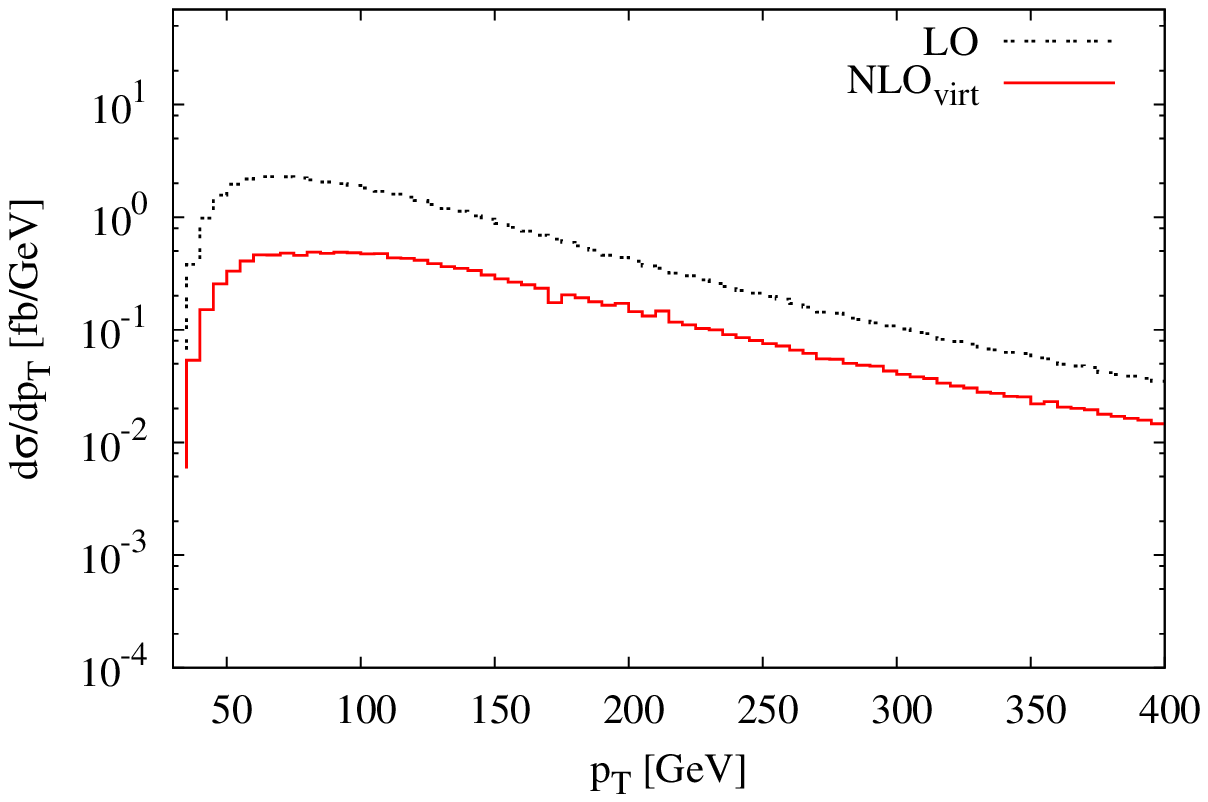}%
\includegraphics[width=0.45\textwidth]{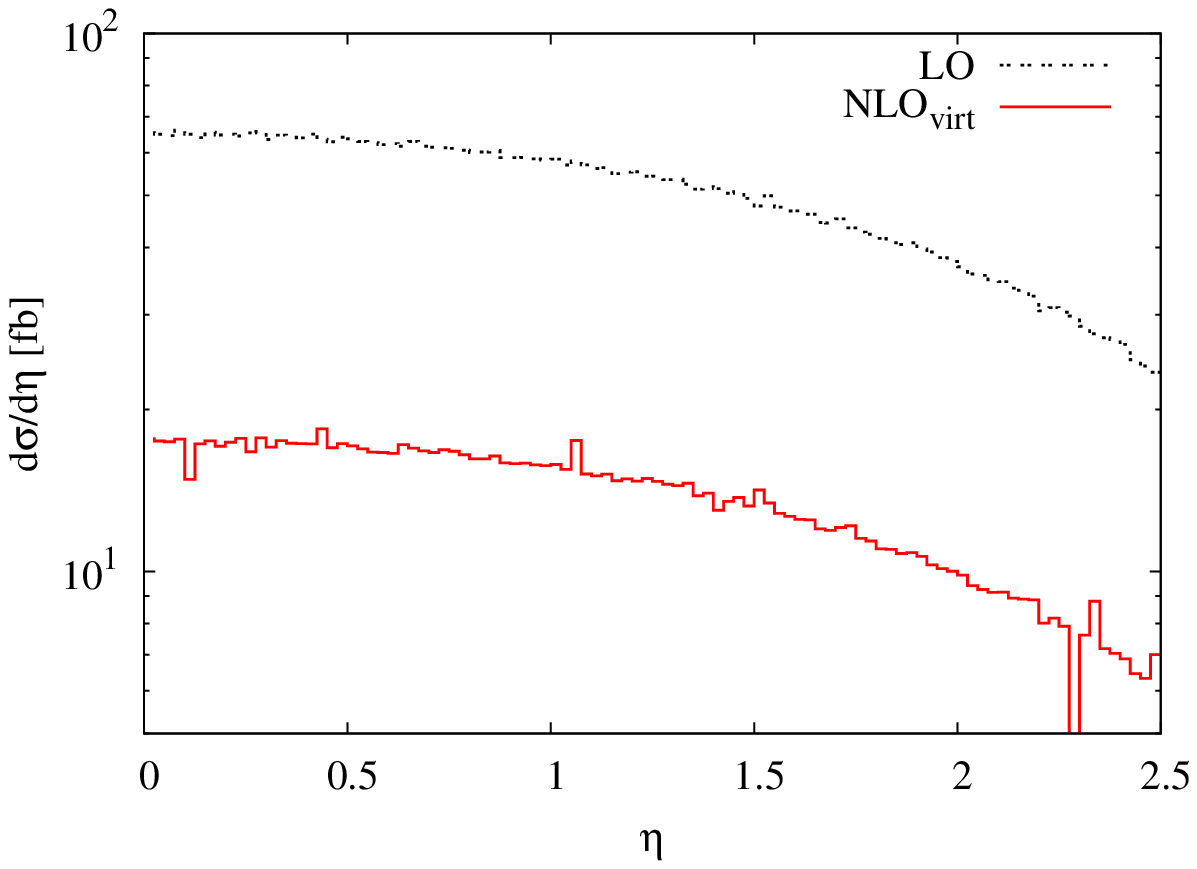}
\end{center}
\caption{Distributions of the transverse momentum (left) and the
rapidity (right) of the hardest jet.
The result for $\mathrm{NLO}_{\mathrm{virt}}$
is obtained from the finite contribution of the virtual part
of the NLO prediction, as described in Eq.~(\ref{eq:local-K}).}
\label{fig:jet-pt}
\end{figure}

\begin{figure}[htbp]
\begin{center}
\includegraphics[width=0.9\textwidth]{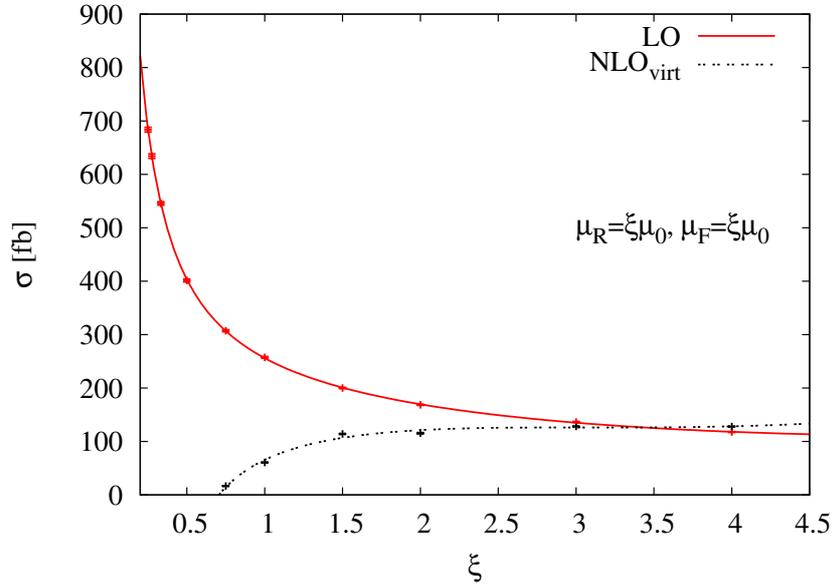}
\end{center}
\caption{Stability under simultaneous variation of the
renormalisation scale $\mu_R$ and the factorisation scale
$\mu_F$ around the central value of $\mu_0=\sum_{i=1}^4p_{T,i}/4$.}
\label{fig:scalevar-par}
\end{figure}

\section{Conclusion}
A successful interpretation of the LHC data will need precise
predictions of both background and signal. For many processes
this involves the evaluation of one-loop QCD amplitudes with
many particles in the final state. We have presented the approach
of the \texttt{GOLEM} collaboration to automatise such calculations
that allows to generate code for the matrix element which is both
fast and numerically stable in all relevant phase space regions.
The need to control the accuracy of the result
has been emphasised and a posteriori precision test has been proposed.
We have introduced a new indirect integration method based on
reweighting unweighted LO events by a local $K$-factor, which avoids
certain problems that otherwise arise from a direct, adaptive MC
integration of the one-loop matrix element.
Results have been presented for the virtual corrections of the
process $u\bar{u}\rightarrow b\bar{b}b\bar{b}$, which is 
part of an important background for MSSM Higgs boson searches at the LHC.

We have shown that our approach allows for efficient implementations
for NLO predictions of processes with multi-particle final states
and that the \texttt{GOLEM} project can lead to a fully automated
tool for NLO calculations.

\section*{Acknowledgement}
The author is grateful to Thomas Binoth,
Alberto Guffanti, Jean-Philippe Guillet and J\"urgen Reuter
for their special support in the $u\bar{u}\rightarrow b\bar{b}b\bar{b}$
project.
He also wants to thank the organisers of the 5th Vienna Central
European Seminar on Particle Physics and Quantum Field Theory for
their great hospitality.  
This work has made use of the resources provided by the Edinburgh
Compute and Data Facility (ECDF)~\cite{ECDF}.
The ECDF is partially supported by the eDIKT initiative\footnote{%
http://www.edikt.org}.

\end{document}